**Aprendizaje de las ecuaciones diferenciales de orden superior través del estudio del movimiento del péndulo simple en el marco de la alianza Internacional World Pendulum Alliance WP@elab**

**Learning of second order of differential equations through the study of the movement of simple within the framework international World Pendulum Alliance WP@elab**

Declaración: el material es original e inédito y no se encuentra en proceso de revisión para publicación.

Dayana Alejandra Barrera Buitrago
Universidad Nacional Abierta y a Distancia Bogotá, Colombia
Escuela de Ciencias Básicas, Tecnología e Ingeniería
dayana.barrera@unad.edu.co
+573002622233
Dirección postal: Av Ciudad de Quito #11 Sur-2 a 11 Sur-48, Bogotá
https://orcid.org/0000-0001-8867-9705

Nidia Danigza Lugo López
Universidad Nacional Abierta y a Distancia Bogotá, Colombia
Escuela de Ciencias Básicas, Tecnología e Ingeniería
nidia.lugo@unad.edu.co
+573123675024
Dirección postal: Av Ciudad de Quito #11 Sur-2 a 11 Sur-48, Bogotá
https://orcid.org/0000-0002-9096-5767

Freddy Alexander Torres Payoma
Universidad Nacional Abierta y a Distancia Bogotá, Colombia
Escuela de Ciencias Básicas, Tecnología e Ingeniería
freddy.torres@unad.edu.co
+57 3005141222
Dirección postal: Av Ciudad de Quito #11 Sur-2 a 11 Sur-48, Bogotá
https://orcid.org/0000-0002-5206-0836

Diana Carolina Herrera Muñoz
Universidad Nacional Abierta y a Distancia Bogotá, Colombia
Escuela de Ciencias Básicas, Tecnología e Ingeniería
diana.herrera@unad.edu.co
+57 321 9753048
Dirección postal: Cl. 14 Sur # 14-23, Bogotá
https://orcid.org/0000-0003-4188-3412

Laura Daniela Neira Quintero
Universidad Nacional Abierta y a Distancia Bogotá, Colombia
Escuela de Ciencias Básicas, Tecnología e Ingeniería
ldneiraq@unadvirtual.edu.co
Dirección postal: +57311 3835880
Av Ciudad de Quito #11 Sur-2 a 11 Sur-48, Bogotá
https://orcid.org/0000-0002-4304-4922

**Resumen**

Esta investigación se trata de analizar los efectos de la implementación de un espacio virtual de aprendizaje de las ecuaciones diferenciales de segundo orden homogéneas mediante la modelación del fenómeno físico del péndulo simple en el marco de la alianza con el proyecto internacional World Pendulum Alliance. La recolección de la información se realizó con un cuestionario de cinco preguntas, usando pretest y postest. La efectividad de la estrategia se midió con la ganancia de Hake y la evolución de los aprendizajes con los vectores de Bao. Se encontró una efectividad mayor pero cercana a la que logran los modelos tradicionales, por lo que los estudiantes lograron los objetivos de aprendizaje propuestos. La evolución del aprendizaje sugiere que los estudiantes posteriores a la implementación de la estrategia se enmarcan bajo el modelo de explicación aceptado en la literatura, aunque algunos participantes todavía se inclinan por el modelo incorrecto o el distractor. Por ende, es importante incluir en la estrategia más actividades relacionadas a la descripción del fenómeno físico y en la mejora de los procedimientos algebraicos vistos en cursos previos.

**Palabras clave:** Enseñanza de las matemáticas, educación a distancia, modelo matemático, matemáticas, física.

**Abstract**

This research is about to analyze the effects of the implementation of a virtual learning space in the homogeneous second-order differential equations applied to the modeling of the phenomenon of simple pendulum physics in the framework of the alliance with the international project World Pendulum Alliance WPA. The collection of information was carried out with a five-question, using pre-test and post-test. The effectiveness of the strategy was measured with the Hake gain and the evolution of learning with the vectors of Bao. It was found that the strategy is more effective, but close to that achieved by traditional models, so that students largely achieve the proposed learning objectives. The evolution of learning suggests that students after the implementation of the strategy are marked under the model of acceptable explanation in the literature, although some participants still lean towards the incorrect model or the distracter. Therefore, it is necessary to include in the strategy a greater number of activities related to the description of the physical phenomenon to be studied and the improvement of the algebraic procedures seen in previous courses.

**Key words:** Mathematics teaching, e-learning, mathematical model, mathematics, physics.

# 1. Introducción

La Universidad Nacional Abierta y a Distancia (UNAD) es una institución de educación superior con modalidad abierta y a distancia que tiene como principal misión la educación para todos, mediante: la acción pedagógica, la proyección social, el desarrollo regional, la acción comunitaria, la inclusión, la solidaridad, la investigación, la internacionalización y la innovación en todas sus expresiones; para fomentar el aprendizaje autónomo, significativo y colaborativo (Universidad Nacional Abierta y/a Distancia, 2021). En el marco de los Objetivos de Desarrollo Sostenible (ODS) (Oficina Internacional del Trabajo - Alianza Cooperativa Internacional, 2014; Organización de las Naciones Unidas, 2014) la UNAD está comprometida con el ODS 4 "Educación de Calidad", garantizando la inclusión, equidad y calidad con su política social, siendo la única universidad pública de Colombia que acepta a cualquier ciudadano con el deseo de estudiar que esté en posesión de un diploma de educación media o bachillerato, sin importar la zona del país en la que se encuentre o las necesidades especiales de los estudiantes, sin realizar ningún procedimiento de ingreso. En ese sentido, los estudiantes de la UNAD por ser una población diversa a nivel nacional, en cuanto a su condición social, cultural y económica; hace que, dentro del contexto de la universidad, se presenten retos especiales para el aprendizaje, dando cumplimiento de los criterios de inclusión social educativa, fomentados en el Proyecto Académico Pedagógico Solidario (PAPS) faciliten la vinculación de poblaciones de estudiantes situados en diversas partes del mundo (UNAD, 2011).

Los estudiantes en general coinciden en tener dificultades en el aprendizaje de las Ecuaciones Diferenciales (ED) tanto de forma conceptual, como procedimental (Guerrero, 2010; Arslan, 2010). En general, en diferentes trabajos académicos que han mostrado mejoras en el aprendizaje de las ED a través de la modelación matemática y el uso de tecnología (Rasmussen y King, 2000; Chaachoua y Saglam, 2006; Ortigoza, 2007; Rodriguez, 2010; Carranza, 2018; Castrillo, 2022). Una manera de tratar de mitigar las dificultades académicas presentadas por los estudiantes en el aprendizaje de las ED es hacer que ellos conozcan su origen a partir del modelado de fenómenos físicos, que es donde se han identificado dificultades en los procesos de enseñanza-aprendizaje bajo los conceptos matemáticos que se exponen los libros de texto en su contenido, los cuales omiten la relación histórica directa con la física desde sus orígenes en el Siglo XXI (Henao y Rodríguez, 2016).

El desarrollo de la presente investigación subyace en el marco del proyecto World Pendulum Alliance WP@ELAB cofinanciado por el programa Erasmus+ de la Unión Europea (EU) mediante el convenio de subvención 598925-EPP-1-2018-PT-EPPKA2-CBHE-JP (Instituto Superior Técnico ITS, 2018). El proyecto es coordinado por el Instituto Superior Técnico – Universidad de Lisboa IST y tiene como objetivo establecer una red de péndulos con acceso remoto, ubicada en diferentes puntos geográficos. El acceso a cada experimento se realiza a través de internet de manera sincrónica y de una aplicación que permite transmitir y visualizar los datos obtenidos en tiempo real. El objetivo principal del proyecto es mejorar la calidad de la educación media y superior, en campos como las matemáticas y las ciencias naturales. Entre los objetivos específicos del proyecto es la creación de cursos masivos, abiertos y en línea que permitan dar un mejor aprovechamiento a la red pendular o que aporten para la educación en matemáticas, que es justo donde el proyecto apoyó esta investigación dando capacitación docente en la creación de espacios virtuales de aprendizaje, específicamente en el recurso pedagógico de acceso libre Graasp de Ecuaciones diferenciales publicado en la página https://graasp.eu/s/sjcuja.

Go-Lab es un proyecto financiado por la Unión Europea (UE) dentro del Programa FP7 durante cuatro años desde 2013 a 2016 (Jong et al., 2021), que tiene como objetivo principal la mejora de la enseñanza STEM en las aulas desde la perspectiva de la indagación científica (Menchacha, 2020). Además, Go-Lab está constituido por dos sistemas en línea: el primero es el Sistema de Soporte y Compartición llamado Golabz, este es el encargado de incorporar los recursos digitales y el segundo es el Sistema de Autoría (Graasp) un constructor ILS gratuito, el cual cuenta con herramientas y recursos virtuales para la elaboración de prácticas a distancia (Heradio et al., 2016) y se fundamenta en el aprendizaje basado en indigación (ABI), simulaciones en laboratorios y entornos simulados de aprendizaje(Jong, Lazonder, Pedaste y Zacharia, 2018). El Graasp es una plataforma de aprendizaje social con un sistema de autoría que fundamenta el aprendizaje, investigación, uso de laboratorios en línea y creación de espacios personales (Gillet, Vozniuk, Rodriguez-Triana y Holzer, 2016), entre tanto, Golabz es el sistema que permite el uso de recursos interactivos que apoyan la educación STEM. Además de fomentar la creación e implementación de laboratorios para enriquecer la experiencia formativa con la participación colectiva y eventualmente su difusión a través de Creative Commons, con un enfoque en brindar acceso libre a recursos en línea (Palau, Mogas-Recalde y Domínguez-García, 2020). El Go-Lab fusiona diferentes laboratorios online con ILS y otros recursos de aprendizaje como herramientas interactivas donde el sistema de creación y edición permite a los profesores desarrollar entornos de aprendizaje originales.

En consecuencia, esta investigación tiene como objetivo general analizar los resultados obtenidos en la implementación de un espacio virtual de aprendizaje de las ecuaciones diferenciales a través del entorno Graasp, con la finalidad de ir más allá de los aprendizajes netamente memorísticos para la solución de las ED y llevar a los estudiantes a la comprensión y la interpretación de los conceptos que se relacionan naturalmente con el fenómeno físico del péndulo simple mediante la implementación de espacios virtuales de aprendizaje basado en indagación ABI (Palau, Mogas-Recalde y Domínguez-García, 2020). Por lo anterior, en este artículo se presenta el contenido educativo desarrollado en la plataforma Graasp como herramienta de ayuda en la enseñanza de ecuaciones diferenciales en el curso impartido por parte de la Universidad Nacional a Distancia y Abierta. Este curso está orientado a programas de ingeniería: industrial, sistemas, telecomunicaciones, electrónica, alimentos, medio ambiente y agroforestería, es un componente de la formación en ciencias básicas. A continuación, se describe la estructura general del curso de ED.

El syllabus del curso de ED consta de tres unidades temáticas: *ecuaciones diferenciales de primer grado; ecuaciones diferenciales de orden superior;* y *transformadas de Laplace*, el lector podrá encontrar en (Leal, 2021) la consolidación de la metodología de enseñanza abierta y a distancia de la UNAD basado en Aprendizaje Basado en Tareas

Aprendizaje basado en indagación (ABI)
 Aprendizaje Basado en Tareas (ABT)

(ABT). La implementación del Go-Labz de ED se emplea, específicamente en la Unidad 2, en el tema de "Ecuaciones Diferenciales Homogéneas de Segundo Orden. El objetivo de aprendizaje de la implementación del Graasp en abordar es: "*Solucionar con diferentes métodos matemáticos situaciones problema que involucran las ecuaciones diferenciales de orden superior, con el fin de dar solución a problemas relacionados con la ciencia e ingeniería*" El espacio de Aprendizaje por Indagación (ILS de sus siglas en inglés) desarrollado en Graasp Ecuaciones Diferenciales posee seis escenarios de aprendizaje y dos criterios de acceso a datos y escenarios de control para estudiante y profesor. El espacio está alojado en la página ww.graasp.edu, el lector podrá ingresar y hacer uso del espacio de forma libre siguiendo el enlace https://graasp.eu/s/sjcuja. Los contenidos desarrollados en el ILS se explican en la Tabla 1.

*Tabla 1.*
*Escenarios del Graasp Ecuaciones Diferenciales*

| Tópico | Descripción |
|---|---|
| Introducción (Orientación) | Se registran los temas y objetivos del recurso pedagógico. |
| Evaluación Inicial (Orientación) | Cuestionario PRE (Prueba Objetiva Cerrada Inicial) para evaluar los presaberes. |
| Contenido (Conceptualización) | Contenido pedagógico de las ecuaciones diferenciales que modelan el fenómeno físico del péndulo simple. |
| Actividad de aprendizaje (Investigación) | Contenido pedagógico de las ecuaciones diferenciales que modelan el fenómeno físico del péndulo simple |
| Evaluación Final (Conclusión) | Cuestionario POS (Prueba Objetiva Cerrada Final) |
| Conclusiones (Discusiones) | En este espacio, se evalúan las discusiones y conclusiones al finalizar la actividad. |

La construcción del espacio ILS se basa en la indagación orientada, donde el docente tendrá la función de guiar a los estudiantes a través de encuentros sincrónicos para lograr los objetivos de aprendizaje. (Liang, Deng, y Liu, 2008; Sbarbati Nudelman, 2015). Cabe resaltar que los métodos de cuestionario Pre y Pos son del tipo Prueba Cerrada Objetiva, los instrumentos serán descritos a profundidad en la sección de Método.

Este trabajo se organiza de la siguiente manera: primero se presenta el marco teórico donde se describe el fundamento físico, la estrategia de aprendizaje y la forma como se estudiará la evolución del aprendizaje; posteriormente la sección de métodos, donde se explica la implementación pedagógica y las metodologías que se usan para analizar los resultados obtenidos por los estudiantes ; continuando con los resultados y discusión; y finalizando con las conclusiones.

## 2. Marco teórico

En esta sección se desarrollan los elementos teóricos que son importantes para el trabajo, entre los que se encuentra: laboratorios virtuales y laboratorios remotos, aprendizaje por indagación, estos tres últimos usados por el Graasp, herramienta empleada para el desarrollo de la investigación, seguido se presenta la ecuación de movimiento para el péndulo simple, aprendizaje basado en tareas, aprendizaje basado en indagación, la ganancia de Hake y los índices de concentración Bao para los modelos alternativos que los estudiantes emplean para la explicación de los fenómenos

*2.1. Aprendizaje basado en Indagación*

El Aprendizaje Basado en Indagación (ABI) es una estrategia educativa basada en métodos y prácticas profesionales, con el objetivo de indagar en el conocimiento científico (Keselman, 2003). El enfoque de la estrategia está orientado en la formulación de una hipótesis que plantee un problema y logre validar el resultado a través del desarrollo de experimentos. El ABI enfatiza la participación permanente de los estudiantes, resaltando las responsabilidades que debe adquirir en el momento de formular una hipótesis y explorando su respuesta mediante experimentos (Pedaste, y otros, 2015).

La propuesta ABI implementada en aula debe cumplir cinco fases para evaluar, de forma adecuada, el proceso de enseñanza-aprendizaje, las cuales son: *orientación, conceptualización, investigación, conclusión y discusión*. En la figura 1 se explican las fases y subfases del ABI, definiciones y características principales. El texto de Pedaste, et al. (2015) muestra una revisión completa sobre la estructura del ABI.

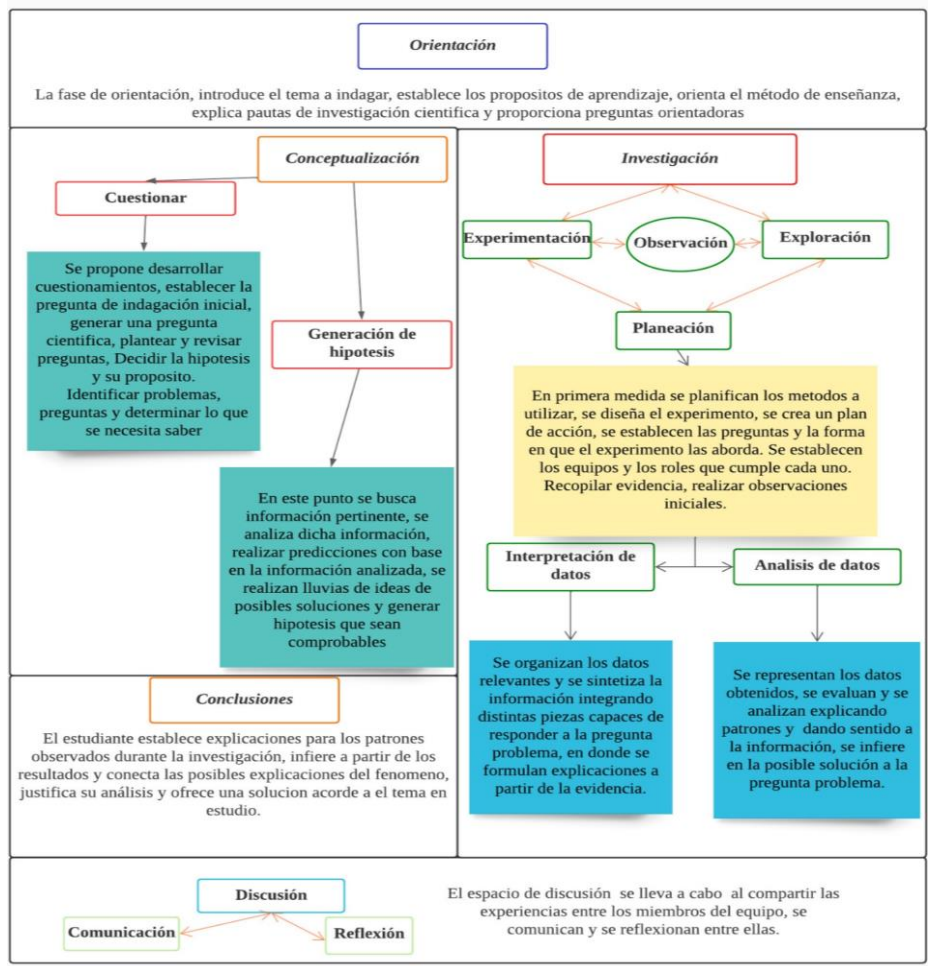

*Figura 1. Fases del Aprendizaje Basado en Indagación. Adaptado (Pedaste et al., 2015).*

## 2.3. Estructura de las cinco fases del ABI.

Los Espacios Virtuales de Aprendizaje (ILS de sus siglas en inglés), son lugares dedicados en la Web para construir prácticas de laboratorio basadas en la estructura ABI (Araque, Montilla, Meleán, y Arrieta, 2018)

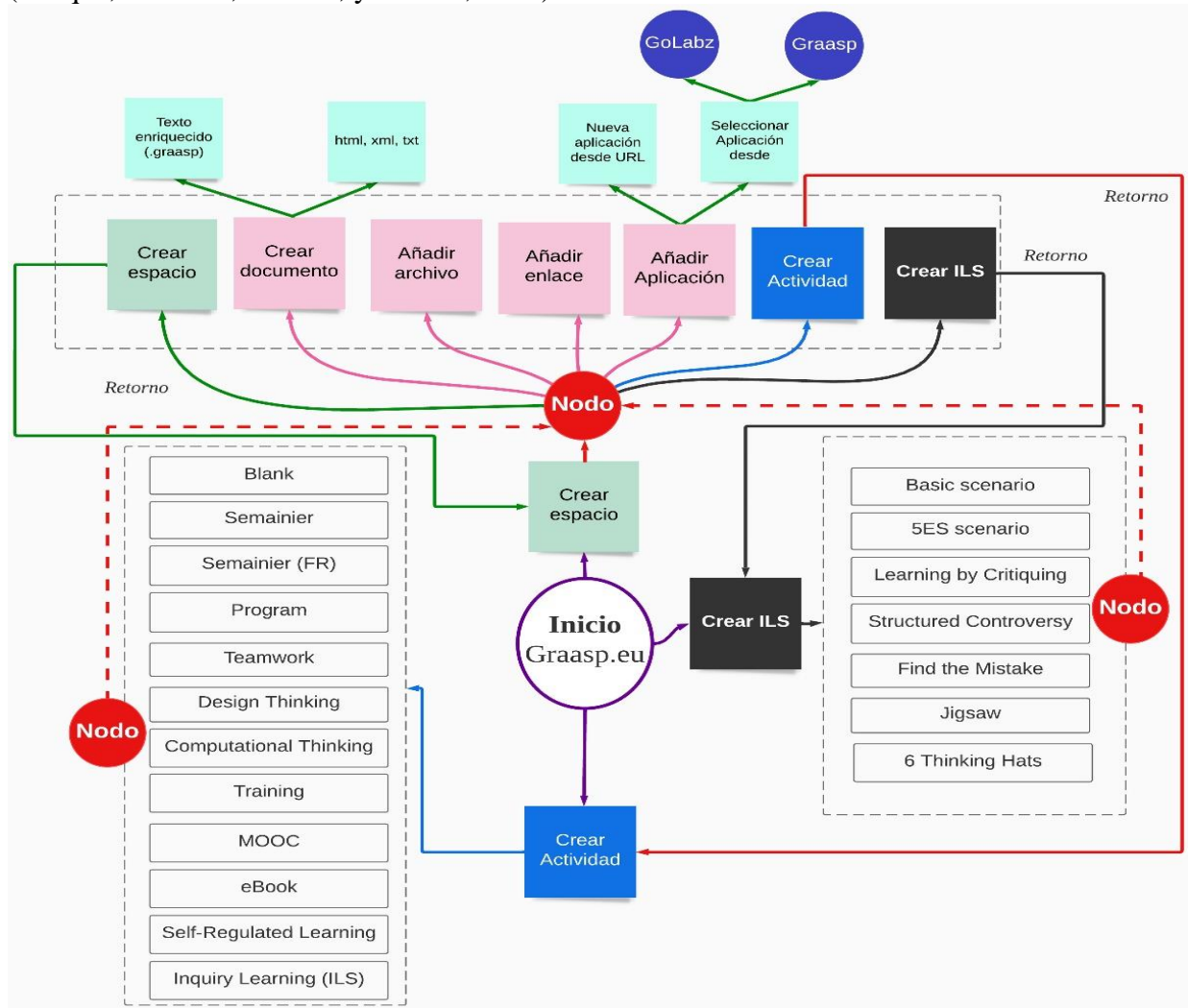

*Figura 2. Estructura generalizada de la construcción de un Go-Lab a partir de Graasp. Adaptado (Torres-Payoma, F., Herrera, D. C., Tique, D. L., Barrera, D., y Penalba, F. A, 2022).*

El Proyecto Go-Lab de Next-Lab utiliza los laboratorios remotos y virtuales para promover la enseñanza de las ciencias mediante la interacción de un ecosistema guiado y orientado al aprendizaje por indagación (Palau, Mogas-Recalde, y Domínguez-García, 2020) .

La estructura general del ILS basado en ABI en la plataforma Graasp se presenta en la Figura 2, en (Torres-Payoma, F., Herrera, D. C., Tique, D. L., Barrera, D., y Penalba, F. A, 2022) se realiza la creación y ejemplo de cada uno de los pasos para la construcción de un ILS de Conservación de la Energía y su implementación en el aula virtual.

## 2.4. Péndulo Simple

El péndulo simple o péndulo matemático es un sistema físico ideal que es utilizado para introducción al estudiante en el estudio de las vibraciones y ondas, adicionalmente es empleado en la enseñanza de las ecuaciones diferenciales como un problema de aplicación en contexto reales. Este consiste en una partícula de masa m suspendida con una cuerda ligera de longitud $L$, fija en el extremo superior como se muestra en la Figura 3.

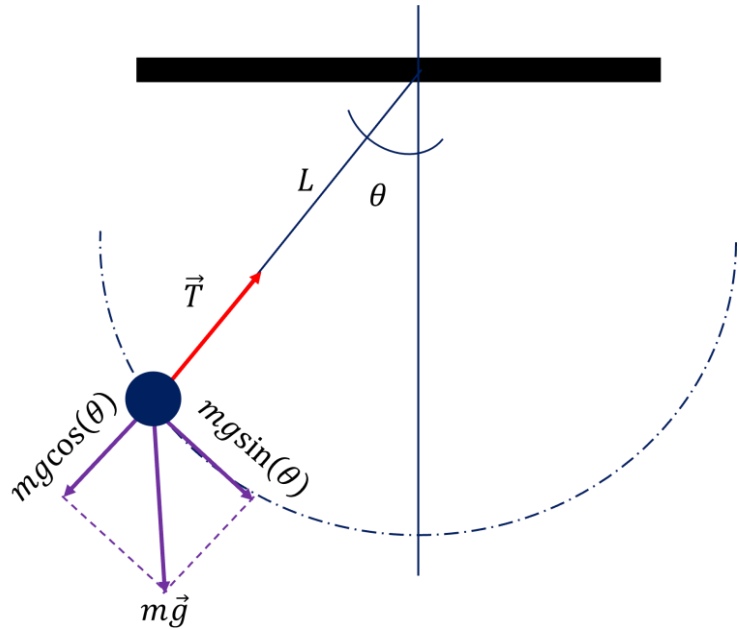

*Figura 3. Péndulo simple. Adaptada de Serway y Jewett (2009).*

Sobre la masa $m$ actúan la tensión $\overline{T}$ debida a la Cuerda y la fuerza gravitacional $m\overline{g}$. Si se realiza la suma de fuerzas sobre la componente tangencial se puede aplicar la segunda ley de Newton sobre esta misma dirección:

$$F_T = -mg\, sen\, \theta = \frac{d^2 s}{dt^2} \quad (1)$$

El signo menos indica que esta componente del peso es una fuerza restauradora, s es la posición de la plomada medida a lo largo del arco. Ahora recordando que $s = L\theta$, donde L es la longitud de la cuerda. Reemplazando en la ecuación anterior se encuentra la siguiente ecuación diferencial.

$$\frac{d^2\theta}{dt^2} + \frac{g}{L} sen\, \theta = 0 \quad (2)$$

Teniendo en cuenta que $\theta$ indica la forma en cómo se mueve la masa m respecto a la posición de equilibrio, esta ecuación es similar a la expresión general para el Movimiento Armónico Simple (MAS), la diferencia radica es que la posición cambia en función del seno del ángulo en lugar de $\theta$, por lo tanto, para considerar al movimiento del péndulo un MAS es necesario tomar valores del $\theta$ para los que el $sen\,\theta \approx \theta$, esto es, ángulos pequeños que como señala Serway y Jewett (2009) deben ser menores a 10º. De modo que, para ángulos pequeños la ecuación de movimiento se convierte en:

$$\frac{d^2\theta}{dt^2} + \frac{g}{L}\,\theta = 0 \tag{3}$$

La cual es una ecuación diferencial de segundo orden, cuya solución es:

$$\theta = \theta_0 \cos(\omega t + \phi) \tag{4}$$

Esta situación fue propuesta como problema de aplicación en el curso de ecuaciones diferenciales de la universidad UNAD. Para esto se les propuso a los estudiantes un problema de aplicación que tenia de base un péndulo simple, ellos debían plantear la ecuaciónde movimiento y solucionarla.

## 2.5. Estrategia de enseñanza empleada en el aula de clase

Con el propósito de describir la estrategia empleada en el curso de ecuaciones diferenciales se debe partir por explicar grosso modo el Modelo Pedagógico (MP) que se emplea en la UNAD y la estrategia de aprendizaje que se emplea en muchos de sus cursos. El MP Unadista tiene por lema el Aprender a Aprender y para que esto se logre se debe consolidar un conjunto estructuras que conduzcan a este camino, además se deben abandonar los modelos convencionales de enseñanza. Los actores principales del proceso, docente y estudiante dejan atrás la postura tradicional, el primero toma un papel de líder, orientador o tutor, mientras que el segundo se hace activo, auto formativo, autónomo y logra trabajar de manera colaborativa, todo esto acompañado de e-medios, e-mediaciones, e-evaluaciones, e-comunicación y e-investigación (Leal Afanador, 2021).

Adicional a lo anterior, se proponen escenarios y estrategias para el acompañamiento docente que se emplearon de manera regular en el aula de ecuaciones diferenciales. Las webconferencias: espacios de atención sincrónica que se pueden dar de manera presencial o virtual y se enfoca en describir las generalidades del curso y desarrollar los contenidos programáticos del mismo y los Círculos de Interacción y Participación Académica y Social (CIPAS) que se puede definir como "comunidades de aprendizaje, conformadas entre estudiantes, y orientadas por un docente. En ellas se dan interacciones estudiante-estudiante y estudiante-docente, para resolver inquietudes entre pares sobre el aprendizaje, el desarrollo de los cursos y programas académicos" (Leal Afanador, 2021, p. 78).

Por último, se debe mencionar que el curso empleó como estrategia de aprendizaje el Aprendizaje Basado en Tareas (ABT), definido como un aprendizaje con enfoque por tareas, que logra transformar la enseñanza tradicional basada en el profesor a una enseñanza basada en el estudiante, puesto que en el ABT el estudiante va ganando de forma progresiva responsabilidad con su aprendizaje a partir de la solución de problemas propios de su profesión lo que le da motivación y permite un aprendizaje significativo (Liang, Deng y Liu, 2008).

Una vez que se han expuesto las generalizadas del MP de la UNAD y las formas de acompañamiento docente que se utilizan, se mostraran las generalizadas del curso. Ecuaciones diferenciales es una materia obligatoria de componente de ciencias básicas de las ingenierías, tiene tres créditos académicos, que se divide en tres unidades:

- Unidad 1. Ecuaciones diferenciales de primer orden.
- Unidad 2. Ecuaciones diferenciales de orden superior.
- Unidad 3. Solución de ecuaciones diferenciales mediante transformada de Laplace.

Esta investigación se desarrolló en la unidad 2 que contempla las siguientes subunidades temáticas:

- Tema 1: Ecuaciones Diferenciales Homogéneas.
- Tema 2: Ecuaciones Diferenciales No Homogéneas.
- Tema 3: Ecuaciones Diferenciales Cauchy - Euler
- Tema 4: Aplicaciones de las ecuaciones diferenciales de orden superior.

Estos cinco temas, los estudiantes los trabajan dentro de una sola tarea que tiene una duración de cuatro semanas, uno por cada tema, donde la actividad del Graasp formó parte del ejercicio que trata el tema 4, teniendo un valor de 40 puntos sobre 100 puntos posibles. Para su implementación directamente con los estudiantes se elaboró un Grassp. El resultado del aprendizaje de la unidad 2 al que responde esta estrategia de enseñanza es: Solucionar con diferentes métodos matemáticos ecuaciones diferenciales de orden superior que involucran situaciones problema relacionadas con la ciencia e ingeniería, lo que justifica que en el recurso educativo digital Graasp, se presente la aplicación de las ecuaciones diferenciales de segundo orden homogéneas en un problema relacionado con la ciencia específicamente el estudio del péndulo simple.

Para ayudar a guiar a los estudiantes hacia un correcto uso del Graasp, los docentes realizaron clases grabadas, web conferencias y CIPAS y web con videotutoriales con el fin de que sincrónica o asincrónicamente presentar el contenido programático de mismo, descrito a continuación:

- ✓ El resultado del aprendizaje del Graasp: introducir al estudiante a las ecuaciones diferenciales de segundo orden mediante un modelo físico sencillo, el péndulo simple.
- ✓ Evaluación inicial: cuestionario de 7 preguntas para hacer un diagnóstico de los estudiantes (pretest).
- ✓ Contenido del aprendizaje: es la explicación de la definición del péndulo simple, la ecuación diferencial que modela el péndulo simple, las soluciones de la ecuación diferencial con el método analítico y el de Euler. Se utilizó un ILS Graasp que permite interactuar con laboratorios interactivos donde se explica el fenómeno físico del péndulo simple, como se refleja en la Figura 4.
- ✓ Actividad del aprendizaje: se propuso una actividad para solucionar situaciones problemas planteadas con los métodos analítico y Euler.

- ✓ Evaluación final: es el mismo cuestionario del pretest, pero los estudiantes lo realizaron después de seguir los anteriores pasos.
- ✓ Conclusiones: se concluye la importancia de los problemas simples de la física en la compresión de las ecuaciones diferenciales.

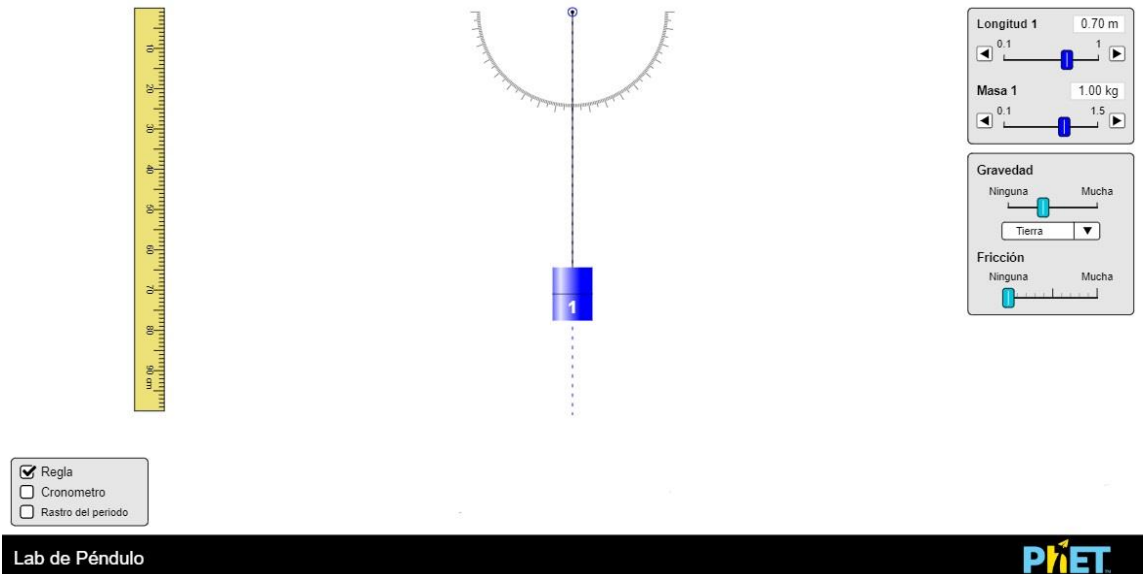

*Figura 4. Laboratorio de péndulo simulador del movimiento que se encuentra en el contenido del Graasp.* Fuente: Universidad de Colorado Boulder (2022).

La implementación del Graasp con los estudiantes tuvo los siguientes pasos:

1. La actividad se encuentra planificada en la malla curricular del curso y por lo que se especifica en la agenda al inicio del semestre.
2. Durante la actividad se realizaron dos sesiones virtuales (web conferencias) para explicar el contenido del recurso educativo digital del Graasp, que consiste en una página web conde se explica los temas del péndulo simple, su ecuación diferencial, la forma de solucionar situaciones problema y el método de Euler. Se tuvo la participación sincrónica de 25 estudiantes.
3. En la primera web conferencia se realizó el pretest y la explicación de los contenidos relacionados con el péndulo simple, su relación con las ecuaciones diferenciales y la solución analítica.
4. En la web conferencia se explicó la solución de la ecuación diferencial con el método de Euler, también cómo se debía entregar la actividad del aprendizaje, se resolvieron preguntas y se puntualizó en indicar que el postest al final de las actividades previas.
5. Como material adicional se realizaron dos tutoriales donde se explicó paso a paso la realización de la actividad del aprendizaje, debido a que en la sesión 2, quedaron dudas pendientes.
6. Los tutores realizaron una web conferencia y dos CIPAS para aclarar dudas presentes. Además de enviar mensajes recordatorios del cierre de la actividad.
7. Se les indicó a los estudiantes que cuando terminaran de realizar todos los ejercicios de la unidad 2, realizaran el postest del Graasp.

## 2.6. Observables de monitoreo de la evolución del aprendizaje

Con el propósito, de determinar el alcance de la estrategia pedagógica implementada en el curso de ecuaciones diferenciales, en esta investigación se hace uso de la ganancia de Hake (g) y los factores de concentración de Bao (C). El primero es un parámetro estadístico que permite dar información acerca de la evolución del aprendizaje de un grupo de estudiantes a los que se les aplica una estrategia de enseñanza, sin importar el estado de los conocimientos previos de cada individuo. Al mismo tiempo, permite comprobar si una metodología de enseñanza es positiva, considerando el conocimiento previo del estudiante. Su cálculo se realiza con la siguiente ecuación:

$$\langle g \rangle = \frac{\langle S_f \rangle - \langle S_o \rangle}{1 - \langle S_o \rangle}$$

La ganancia de Hake $g$ es baja si esta no supera el 0.3, es media si está entre 0.3 y 0.7 y es alta si es mayor a 0.7. Con el fin de profundizar en mayor medida en la evolución conceptual de los estudiantes, se usan los vectores de evolución de Bao los cuales se construyen desde el factor de concentración (Bao y Redish, 2001). Para la correcta aplicación de esta técnica es necesario tener una prueba de opción múltiple con única respuesta aplicada a un grupo de alumnos, y estudiar la distribución de las respuestas del test, empleando el número de respuestas correctas e incorrectas, esta últimas usualmente ignoradas en los análisis que se realizan de este tipo de pruebas. Este se determina usando la siguiente fórmula:

$$C = \frac{\sqrt{m}}{\sqrt{m}-1} \left( \frac{\sqrt{\sum_{i=1}^{m} n_i^2}}{N} - \frac{1}{\sqrt{m}} \right)$$

donde m es el número de opciones, N la cantidad de estudiantes al que se le aplicó la prueba, ni es el número de estudiantes que escogieron la respuesta i de una pregunta particular. Esté factor toma valores en el intervalo de [0,1], C=0 indica que la distribución de la respuesta se asemeja más a una distribución aleatoria y C=1 que los estudiantes seleccionaron una única respuesta. Cualquier otra situación toma valores entre 0 y 1 (Bao y Redish, 2001).

Para ampliar el panorama es necesario construir la gráfica "S vs C", en donde el primero se conoce como puntaje, el cual

$$S = \frac{A_i}{A}$$

En palabras de Barbosa (2021) la relación existente entre C y S:

> representa un espacio de puntos de «concentración» y «puntaje» que precisa el estado de comprensión del estudiante para un instante cuando contesta una pregunta que ha desencadenado el uso de sus modelos de razonamiento. Al ir aprendiendo en el tiempo, el punto representativo del estado de comprensión se mueve en el espacio de configuraciones SC describiendo una curva que podría denominarse la curva de aprendizaje. Si solo se examina en dos momentos, tenemos dos parejas de datos para construir tan solo el largo de un segmento

dirigido y que bien podría denominarse el vector de aprendizaje del estudiante. (p. 150)

Este espacio de configuraciones se muestra en la Figura 5, en ella se evidencia que se pueden tener diferentes opciones, que se representan en la gráfica con sus siglas en inglés, esto es puntaje y concentración alto como (HH), puntaje y concentración medios (MM) y así sucesivamente. Este espacio de configuraciones también permite dividir las repuestas en modelos correctos, incorrectos o mezclados y al azar.

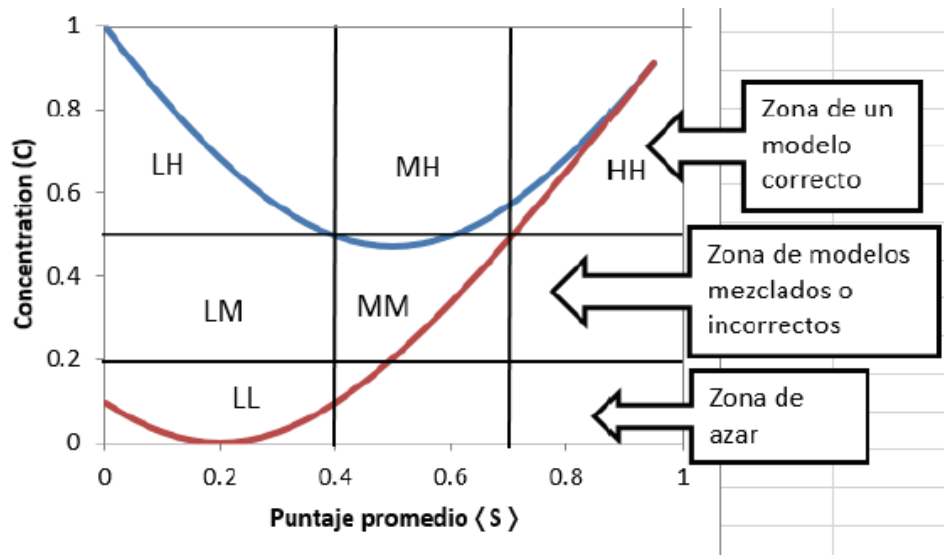

*Figura 5. Gráfica S-C. La línea roja y azul son el límite inferior y superior de la zona de los modelos. Fuente: Tomado de Barbosa (2021).*

Con esta técnica propuesta por Bao (2006) (como se citó en Barbosa, 2021) se puede elaborar el grafico de concentraciones y puntaje por pregunta y de esta manera estudiar el vector de evolución de aprendizaje de los grupos, desde el punto inicial (pretest) hasta el final (postest), este último aplicado posteriormente a la implementación de la estrategia pedagógica.

## 3. Método

Esta investigación se enmarca en el paradigma cuantitativo, bajo una aproximación cuasiexperimental, dado que los estudiantes fueron asignados previamente a los grupos de trabajo, con pretest y postest sin grupo control, para este trabajo se propone el siguiente objetivo: evaluar el impacto que tienen en los estudiantes de la UNAD el aprendizaje de las ecuaciones diferenciales de segundo orden como modelos matemáticos en situaciones problema de la física e ingeniería.

Con esto en mente el pretest y postest fueron empleados para estudiar la evolución de los aprendizajes de la muestra para los cual se utilizó la ganancia de Hake y los factores de concentración, una vez que se implementó a estrategia.

## 3.1. Descripción de la población y muestra intervenida

La investigación se realizó con estudiantes de cuarto semestre de la UNAD de diferentes programas de ingeniería entre los cuales están: alimentos, sistemas, telecomunicaciones, electrónica e industrial que se matricularon al curso de ecuaciones diferenciales, teniendo aprobados los cursos de cálculo diferencial y cálculo integral. La población inicial corresponde a 320 estudiantes, a los que se les envío el cuestionario para que los solucionaran de manera voluntaria, solo 190 respondieron el pretest y de estos 148 estudiantes participaron en el pretest y postest. Por lo que la muestra de partida contaba con 148 alumnos. Con el fin de garantizar la calidad de los datos, se procedió a revisar detenidamente la muestra y eliminar aquellas respuestas que no cumplieran con al menos unos de los siguientes criterios:

- ✓ Si el estudiante presentaba el pretest y el postest el mismo día.
- ✓ Si el estudiante se tomaba menos de 10 minutos para responder el cuestionario.
- ✓ Si el estudiante elegía la misma respuesta en todas las preguntas.

Al aplicar estos criterios de 148 estudiantes, el tamaño de la muestra se quedó con 67.

## 3.2. Validación y Descripción de la Técnica de Recolección e Instrumento

Como técnica de recolección de la información se empleó la encuesta, que como señala Hernández-Sampieri, Fernández Collado y Baptista Lucio (2018) es apropiada cuando se quiere conocer el nivel de conocimientos que un grupo de personas tienen en relación con una temática. En ese orden de ideas el instrumento usado fue un cuestionario. Se construyeron las preguntas de los cuestionarios de tipo conceptual, que permitiera comprender los preconceptos de los estudiantes y al final evaluar el aprendizaje adquirido.

El Instrumento de evaluación del aprendizaje contiene cinco preguntas que se usaron para determinar los presaberes de los estudiantes y posteriormente medir la evolución del aprendizaje.

La validación de este se hizo primero a través de un juicio de expertos, este se compartió con la directora de ecuaciones diferenciales Máster en Estadística Aplicada, dos tutores de la misma materia Máster en Matemáticas y una del curso de cálculo integral, Doctora en Investigación y docencia, ellos hicieron observaciones y sugerencias que se aplicaron al mismo y como resultado se obtuvo la versión final mostrada en los Tabla 2.

*Tabla 2.*
*Enunciados de las preguntas con sus respectivos aprendizajes.*

| Enunciados de las preguntas | Aprendizajes por ítem |
|---|---|
| **Pregunta 1.** Un niño de masa m se está columpiando, se sabe que el movimiento que el niño describe es un Movimiento Armónico Simple (MAS) en donde la variable que cambia es el ángulo. La ecuación de movimiento angular que describe el desplazamiento del niño es una ecuación diferencial de segundo orden que respecto al ángulo. Esta se puede escribir como: <br> A. $\frac{d^3\theta}{dt^3} + \frac{g}{l}\theta = 0$ | El estudiante identifica la ecuación diferencial que modela el péndulo simple. |
| B. $\frac{d^2\theta}{dt^2} + \frac{g}{l}\theta = 0$<br>C. $\frac{d\theta}{dt} + \frac{g}{l}\theta = 0$<br>D. $\frac{d^4\theta}{dt^4} + \frac{g}{l}\theta = 0$<br>Clave B | |
| **Pregunta 2.** Un niño que pesa una masa m se está columpiando. El columpio mide l=1.08m y máximo levanta 10°. El movimiento del columpio describe un movimiento armónico simple con gravedad $g = 9.8\frac{m}{s^2}$ y ángulo $\theta$.<br><br>De acuerdo con los anteriores datos, la solución general $\theta$ de la ecuación diferencial que modela el movimiento en esta situación es:<br>A. $y = c_1 e^{3t} + c_2 e^{-3t}$<br>B. $y = c_1 \cos(3t) + c_2 sen(2t)$<br>C. $y = c_1 e^{3t} + c_2 e^{-2t}$<br>D. $y = c_1 \cos(3t) + c_2 sen(3t)$<br>Clave D | El estudiante soluciona la ecuación diferencial implícita en el modelo físico del péndulo simple. |
| **Pregunta 3.** La ecuación del movimiento de un péndulo simple de longitud $1m$ de su ángulo con respecto al tiempo es $\theta(t) = \sqrt{2}\cos t + \sqrt{2}sent$, esta ecuación se calcula solucionando la ecuación diferencial que modela el movimiento armónico simple para ángulos pequeños y se puede escribir de una forma más sencilla como $\theta(t) = Asen(\omega t + \phi)$, donde $\omega$ es la frecuencia angular, A es la amplitud y $\phi$ es el ángulo de fase. De acuerdo con la anterior información y tomando la gravedad como $g = 9.8\frac{m}{s^2}$, la ecuación del movimiento $\theta(t)$ se describiría:<br>A. $\theta(t) = 2sen(3.13t + 0.7853rad)$<br>B. $\theta(t) = 2sen(3.7t + 0.890rad)$<br>C. $\theta(t) = 2sen(4.15t + 0.5899rad)$<br>D. $\theta(t) = sen(3t + 0.7853rad)$<br>Clave A | El estudiante identifica la ecuación del movimiento armónico simple como la solución de la ecuación diferencial. |
| **Pregunta 4.** La ecuación del movimiento de un péndulo simple del ángulo con respecto al tiempo es $\theta(t) = 0.2654sen(2t + 2.4242rad)$ con la longitud de la cuerda de 2.45m. Esta ecuación se calcula solucionando la ecuación diferencial que modela el péndulo simple para ángulos pequeños. Según esta ecuación al cabo de un segundo el ángulo en el péndulo es:<br>A. -0.2561<br>B. -0.2545<br>C. -0.2570<br>D. -0.2586<br>Clave B | El estudiante emplea la solución particular de la ecuación diferencial que modela el péndulo simple para describir la posición del péndulo en cualquier tiempo t. |
| **Pregunta 5.** Un reloj de péndulo depende del periodo de un péndulo para mantener el tiempo correcto. Suponga que un reloj de péndulo se calibra correctamente a nivel del mar y luego se lleva a lo alto de una montaña muy alta. El reloj ahora se mueve: | El estudiante reconoce el concepto del péndulo en situaciones cotidianas. |

|  |
|---|
| A. Más lento |
| B. Más rápido |
| C. Igual |
| Clave A |

Fuente: elaboración propia.

Posterior al proceso de validación por jueces se procedió a determina la confiabilidad del instrumento, para esto se empleó una Prueba Piloto (PP), herramienta ampliamente utilizada para estudiar la confiabilidad y validez de un instrumento. Con la PP se determinó el índice de confiabilidad empleando el Alpha de Cronbach. Para la muestra de la PP se empleó la técnica propuesta por Viechtbauer et al (2015) (citado por Díaz-Muñoz, 2020), dando como resultado un tamaño de 26 estudiantes, dando un valor de 0.612 basados en el trabajo de Ruiz-Bolívar (2022, citado por Díaz-Muñoz, 2020) la confiabilidad es alta, por lo que se puede decir que el instrumento evalúa la variable que se pretende medir y su nivel de confiabilidad es alto.

## 4. Resultados y discusión

### *4.1. Efectividad de la estrategia didáctica*

Las respuestas de los estudiantes se han obtenido a través de formularios de Google, compilados en una base datos en Excel y se les calculó los porcentajes de los puntajes promedio de las cinco preguntas $S_0$ del pretest y $S_f$ del postest, para posteriormente calcular la ganancia de Hake para cada pregunta, obtienen los resultados en la tabla 2.

Se promedia la ganancia de Hake de todo el instrumento y se encuentra que es de 0.37, que, según Castañeda, Carmona-Ramírez y Mesa (2018) en una ganancia media del aprendizaje, mayor a la señalada por Barbosa (2021) para las estrategias de aprendizaje tradicionales, aunque cercana. Lo que permite concluir que los estudiantes consiguen acercarse a la comprensión del aprendizaje propuesto para la estrategia, aplicación de las ecuaciones a un problema de la física. Resultado similar al reportando por (Suarez, Imbanchi-Rodriguez y Becerra-Rodriguez, 2022) en su investigación acerca de la compresión de los circuitos con mediación de las TIC donde el resultado de la ganancia fue 0.37 encontrándose en intervalo g-media, ellos concluyendo que los participantes consiguen dar cuenta de los conceptos propuestos en la actividad.
Evolución del aprendizaje de los estudiantes.

Usando las respuestas de los estudiantes sobre el cuestionario del Graasp de Ecuaciones Diferenciales en el pretest y postest, se calculan los factores de concentración de Bao, para determinar el impacto del aprendizaje en los participantes. En la gráfica x2 se muestran los datos de concentración (C) en función de puntaje (S). A cada pregunta se le asoció un vector representado como una línea, está comienza en la pareja ordena ($S_i$, $C_i$) (símbolos rellenos) y termina en el punto ($S_f$, $C_f$) (Símbolos vacíos). La longitud e inclinación del vector indican la evolución del aprendizaje de los estudiantes, por ejemplo, la pregunta 1 tiene un vector más largo que las demás, evidenciando una evolución en el aprendizaje mayor.

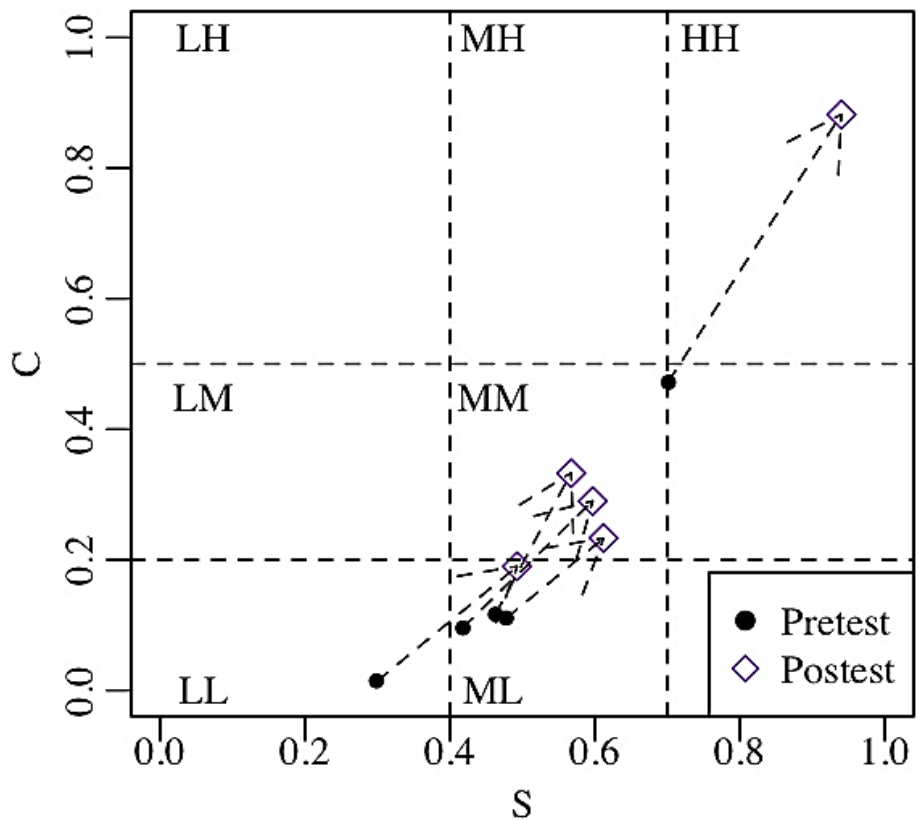

*Figura 6. Datos de vectores evolución de Bao PRE y POS. Elaboración propia. Los puntos negros son los puntos del pretest y los rombos morados del postest. Los puntajes del pretest y el postest se unen con las líneas para cada ítem de la pregunta del cuestionario*

Para el siguiente análisis se toma como referencia la información mostrada en figura 6, al revisar las distribución de las preguntas por zonas se puede observar que en el pretest los datos de puntaje promedio y concentración ubicaron a los estudiantes en las regiones LL (pregunta 3), ML (pregunta 2, 4 y 5), MM (preguntas 1), según la figura 6, siendo LL una región aleatoria, lo que sugiere que los estudiantes seleccionan sus respuestas al azar, sin tener claridad en lo que se les pregunta, algo similar ocurre en la región ML en donde la elección del estudiantado ocurre de la misma forma. Finalmente se cuenta los participantes ubicados en MM en la cual se evidencia la presencia de un modelo correcto y uno incorrecto pero popular.

La información descrita en el párrafo anterior se resume en la Tabla 3, en la cual se muestra el número de la pregunta, el puntaje promedio, la concentración y la zona de ubicación de Bao del pretest (columnas 2, 3 y 4) y postest (columnas 5, 6 y 7).

*Tabla 3.*
*Datos de vectores evolución del aprendizaje para las cinco preguntas.*

| Pregunta | Ganancia de Hake Pretest | Indice de Bao Pretest | Zona Pretest | Ganancia de Hake Postest | Indice de Bao Postest | Zona Postest |
|---|---|---|---|---|---|---|
| 1 | 0.70 | 0.47 | MM | 0.94 | 0.88 | HH |
| 2 | 0.46 | 0.12 | ML | 0.57 | 0.33 | MM |
| 3 | 0.30 | 0.01 | LL | 0.49 | 0.19 | ML |
| 4 | 0.42 | 0.10 | ML | 0.60 | 0.29 | MM |
| 5 | 0.48 | 0.11 | ML | 0.61 | 0.23 | MM |

Fuente: Elaboración propia. En asterisco aparecen los ítems que mostraron cambios en las zonas.

En el postest, las preguntas 1, 2, 4 y presentaron concentraciones y puntajes más altos, en comparación con las otras preguntas (ver tabla 2), por ejemplo, la pregunta 1 en la cual el estudiante debía identificar la ecuación diferencial que modela el péndulo simple, pasó de ser resuelta mayormente por tres modelos; correcto con un 70 %, incorrecto con un 21% y el distractor que no tiene relación directa con la pregunta con un 8%, ubicando esta pregunta en la zona MM a un modelo correcto HH con un 94%.

En el caso de la pregunta 2, en la que se indagaba por la solución de la ED para el péndulo simple, esto es una ecuación de segundo orden, los participantes se repartieron en los mismos tres casos modelo correcto, incorrecto y distractor con porcentaje cercanos 46%, 21 % y 32 %, respectivamente, permitiendo concluir que su selección se debió al azar en el pretest (región ML), mientras que en el postest la mayoría de los participantes emplearon un modelo correcto (57%), la otra parte uso un modelo incorrecto pero popular (34%) y solo un 9% aún se fue por el distractor. Las dificultades presentadas en esta pregunta posiblemente se deban a que se requiere de diferentes cálculos aritméticos o el encontrar la solución a una ecuación cuadrática para encontrar la respuesta correcta frente de la ecuación diferencial que modela el problema del péndulo simple, que se explica según Plaza-Galvez (2016) por la mala formación previa en matemática, es decir, deficiencias en los cursos de matemáticas básicas, cálculo diferencial e integral y álgebra lineal, forma un impedimento para la modelación matemática en el caso de las ecuaciones diferenciales, en ese sentido lo que resulta que pese a las mejoras del aprendizaje en el pretest de la mayoría de los estudiantes, la mala formación previa en matemáticas pudo haber derivado en la selección del modelo incorrecto de algunos estudiantes del curso. Por otra parte, Hernández, Mariño y Penagos (2017) indican que también pueden existir problemas entre el tránsito de la situación problema al lenguaje matemático y al modelo que da solución a la misma, más aún cuando la situación planteada evoca a otro campo de saber cómo es el caso del péndulo situación estudiada en los cursos de física, lo anterior generar que estudiantado necesite conocimientos previos para entender la situación que posiblemente no tenga.

En la pregunta 3, se debe tener en cuenta que las opciones de respuesta solo se dividen en la correcta y las demás erran distractores que proviene del mismo modelo incorrecto, por lo tanto, se evidencia que los estudiantes no logran reescribir la solución de la ecuación diferencial que describe el movimiento de péndulo simple de la forma indicada en la cuestión y que no tiene claridad en cómo hacerlo, entonces no hay un modelo preferencial específico, las 4 opciones tienen porcentajes de respuesta muy cercanos al 25% respondieron de manera aleatoria. Posterior a la implementación, 49 % de los estudiantes aciertan, pero el 51 % restante continúa seleccionado los distractores. Cabe señalar que en la unidad 1 los participantes del estudio vieron cómo solucionar ED de segundo orden y reescribirlas, a pesar de esto ellos no logran hacerlo cuando esta se presenta como solución de una aplicación en el campo de la física, además de tener dificultades en nociones elementales del álgebra sin conseguir cambiar de una ecuación a otra.

Resultados similares son reportados por Hernández et al, (2017) quienes señalan que el alumnado presenta dificultades para transferir lo aprendido de manera algorítmica en la solución ED a problemas de la física, ellos estudian un sistema Masa-Resorte, caso análogo al péndulo simple, y concluyen que el lenguaje matemático y simbólico es complejo para los estudiantes y más si se debe aplicar a una situación problema. Esto concuerda con lo reportado por Hernández-Suarez, Jaimes-Contreras (2016), y Chaves-Escobar (2016), quienes además encuentran que los estudiantes tienen problemas en algunos prerrequisitos de la asignatura "como resolución algebraica, notación y teoremas sobre las derivadas, concepto de antiderivada y teorema fundamental del cálculo" (p. 14).

En el caso de la pregunta 4, las opciones de respuesta son del mismo tipo que la tres y con ella se pretendía evaluar la habilidad del estudiantado de usar la solución de la ED del péndulo simple como una función que describe el movimiento del cuerpo en el tiempo. Por lo que, afirma que antes de la implementación la respuesta de los estudiantes se debió especialmente al azar, ya que no tenían claridad en como emplear la solución, posteriormente, el 60% se agrupan en la respuesta correcta y el 40 % restante en los distractores. De nuevo se evidencia que los participantes tienen problemas comprender el trasfondo físico de la ED, así como, en los procesos algebraico. Resultados similares a los encontrados por Hernández et al, (2017), Hernández-Suarez, Jaimes-Contreras, y Chaves-Escobar (2016) y Plaza-Galvez (2016).

La pregunta 5 está especialmente relacionada con la compresión del fenómeno desde el campo de la física, ya que indaga por los cambios que sufre el periodo al cambiar el valor de la gravedad, con esta mente, los resultados del pretest se dividen dos modelos correcto e incorrecto, dado que solo un 48% señala que disminuye y 52% restante indican que permanece igual o que aumenta. Mientras que, en el postest, el 61% indica la respuesta correcta. De nuevo se puede observar cómo los estudiantes aún tienen dificultades para enlazar el modelo matemático con el físico.

A modo de cierre, se puede decir que las preguntas 2, 3, 4 y 5 tienen algo en común, posterior a la implementación el alumnado aún selecciona el modelo incorrecto pero popular y el distractor como respuesta correcta a la pregunta, en porcentajes mayores al 30% y 15 %, respectivamente. Lo que da evidencia de un avance en la comprensión la ED de segundo orden, dado que logran hacer caso omiso del distractor, cuando la pregunta lo tiene y seleccionar el modelo correcto en la mayoría de los casos o un modelo incorrecto pero popular. Señalando que es necesario trabajar en la estrategia en lo relacionado con mejorar el aprendizaje de los conceptos matemáticos de las EDO de segundo orden cuando se relacionan con el fenómeno físico del péndulo simple y no sólo memorizar los procedimientos para la solucionar las ecuaciones diferenciales como menciona Henao (2022), además de abordar el curso de Ecuaciones diferenciales desde el análisis de situaciones de la física (Hernández et al, 2017).

## 5. Conclusiones

El entorno de aprendizaje del Graasp de Ecuaciones diferenciales enfocado a la aplicación de las EDO de segundo orden homogéneas en el fenómeno físico del péndulo simple mostró mejoraría en la efectividad de la estrategia indicando una ganancia de Hake de

0,37 considerada media y se caracterizó por generar un espacio virtual de aprendizaje que permitió fortalecer los conceptos matemáticos de las ecuaciones diferenciales de orden superior en los estudiantes universitarios de cuarto semestre de ingeniería de la Universidad Nacional Abierta y a Distancia UNAD a través de la modelación en física. El valor de la ganancia de Hake permite concluir que la estrategia diseñada genera una experiencia del concepto de las ecuaciones diferenciales usando aplicaciones de la física y no solo de la exposición de concepto de la ecuación diferencial como una fórmula matemática y una serie de procedimientos para lograr llegar a la solución, carente de todo contexto como se hace usualmente en los cursos tradicionales. Lo que sin duda facilita el proceso de aprendizaje del estudiantado.

Por otra parte, si se asume que la longitud de los vectores de Bao es proporcional a la comprensión que gana el estudiante durante la actividad (Barbosa, 2021), entonces todas las preguntas del cuestionario sobre el tema mostraron una mayor comprensión dado que en el postest las 5 pregunta se movieron a zoma de una mayor compresión, siendo lo más relevante el vector de evolución de la primera pregunta que llegó a la zona HH, lo que quiere decir, que los estudiantes son capaces de identificar la EDO que modela correctamente la situación del péndulo simple. Situaciones similares ocurren con las otras cuatro preguntas con las que se logra que los estudiantes pasen de solucionar el pretest desde el azar, a pesar de tener conocimientos previos sobre cómo solucionar EDO, a agruparse en el modelo correcto con un porcentaje promedio de 57%. Estos resultados también ponen en evidencia que hay características y actividades de la estrategia que se deben mejorar, especialmente las relacionadas con condiciones iniciales, procedimientos algebraicos y descripción más profunda del fenómeno físico a estudiar.

## 6. Financiación

## 7. Referencias